\documentclass{aa}
\usepackage{graphicx}
\usepackage{amsmath}

\begin{document}
\thesaurus{2(12.07.1; 08.02.1; 11.03.1; 11.17.3)}

\title{Secondary caustics in close multiple lenses}
\author{Valerio Bozza\thanks{%
E-mail valboz@sa.infn.it}} \institute{Dipartimento di
Scienze Fisiche E.R. Caianiello,\\
 Universit\`{a} di Salerno, I-84081 Baronissi, Salerno, Italy.\\
 Istituto Nazionale di Fisica Nucleare, sezione di Napoli.}
\date{Received / Accepted }
\maketitle

\begin{abstract}
We investigate the caustic structure of a lens composed by a
discrete number of point--masses, having mutual distances smaller
than the Einstein radius of the total mass of the system. Along
with the main critical curve, it is known that the lens map is
characterized by secondary critical curves producing small
caustics far from the lens system. By exploiting perturbative
methods, we derive the number, the position, the shape, the cusps
and the area of these caustics for an arbitrary number of close
multiple lenses. Very interesting geometries are created in some
particular cases. Finally we review the binary lens case where our
formulae assume a simple form.
\end{abstract}

\section{Introduction}

The binary Schwarzschild lens is one of the most intensively
studied model. In fact, in a relatively simple way, it shows many
features that are observed in general gravitational lenses, such
as the formation of multiple images, giant arcs and a not trivial
critical behaviour. The first study about the binary lens with
equal masses was made by Schneider \& Wei{\ss} (1986). They
derived the critical curves and the caustics showing that three
possible topologies are present depending on the distance between
the two lenses. Erdl \& Schneider (1993) extended these results to
a generic mass ratio of the two lenses. Witt \& Petters (1993)
reached the same results using complex notation. In some limits,
Dominik (1999) enlightened the connection between the caustics of
the binary lens and other models, such as the Chang--Refsdal lens
(Chang \& Refsdal 1979; 1984) and the quadrupole lens.

The critical curves and the caustics of multiple lenses can
develop very complicated structures, so that the attempts to gain
some information about them have been very few. However there is a
great interest in this problem for its applications in particular
situations, such as planetary systems (Gaudi et al. 1998), rich
clusters of galaxies and microlensing of quasars by individual
stars in the haloes of the lensing galaxies (Chang \& Refsdal
1979; Kayser et al. 1988).

In some special situations, the critical curves of multiple lenses
can be derived by perturbative methods, referring to the single
Schwarzschild lens as the starting point for series expansions
(Bozza 1999; Bozza 2000). These methods work very well in
planetary systems, for a lens very far from the others and systems
where mutual distances are very small with respect to the total
Einstein radius. In the first two cases, the complete caustic
structure has been derived and the connections with other models
have been showed. In the last case, only the central caustic
coming up from the deformation of the total Einstein ring has been
studied. Besides this main curve, there are many small critical
curves forming among the masses. The caustics generated by these
curves generally lie far from the centre of mass and can have some
influence on sources distant from the mass distribution. Moreover,
they move very quickly as the parameters of the system change
(Schramm et al. 1993) constituting the most problematic feature to
control in numerical simulations. For these reasons they are
sometimes dubbed \textit{ghost caustics}. In rapidly rotating
binaries they may have superluminal projected motion requiring a non--static
treatment of light deflection (Zheng \& Gould 2000).

In this paper we use complex notation to face the problem of
secondary caustics of close multiple lenses. In this way we can
study them as deeply as the other caustics, completing the previous
works. We shall see that different
classes of secondary caustics can be recognized, showing different
geometries.

After some review of multiple lensing in Sect. 2, in Sect. 3 we
calculate the number and the position of secondary critical curves
for an arbitrary number and configuration of lenses. Then,
in Sect. 4, we treat the
simple caustics and in Sect. 5 the multiple caustics (the
distinction will be explained at the end of Sect. 3). In Sect. 6
we specify our formulae for the binary case and in Sect. 7 we give
the summary.

\section{Basics of multiple lensing}

We shall study a system of n point--lenses placed at positions
$\mathbf{x}_i=\left( x_{i1}; x_{i2} \right)$ in coordinates
normalized to the Einstein radius
\begin{equation}
R_{\mathrm{E}}^{\mathrm{0}}=\sqrt{\frac{4GM_0 }{c^{2}}
\frac{D_{\mathrm{LS}}D_{\mathrm{OL}}}{D_{\mathrm{OS}}}},
\end{equation}
where $M_0$ is a reference mass (it can be chosen to be the total
mass, the typical mass of a single object or anything else). The
source coordinates $\mathbf{y}=\left( y_1; y_2 \right)$ are
normalized to the scaled Einstein radius $R_{\mathrm{E}}^0
\frac{D_{\mathrm{OS}}}{D_{\mathrm{OL}}}$. The masses $m_i$ of the
lenses are measured in terms of $M_0$.

We introduce the complex coordinate in the lens plane $z=x_1+i
x_2$ and the complex source coordinate $y=y_1+i y_2$. The
positions of the masses will be denoted by $z_i=x_{i1}+i x_{i2}$.
We also introduce the functions
\begin{equation}
S_k \left(z \right)=\sum\limits_{i=1}^n \frac{m_i}{\left(z-z_i
\right)^k}. %
\label{Sk}
\end{equation}

The lens equation for our system of n masses reads (Witt 1990)
\begin{equation}
y=z-\overline{S}_1 \left(\overline{z} \right) .%
\label{Lens equation}
\end{equation}
Given a source at position $y$, the $z$'s solving this equation
are the images produced by gravitational lensing.

This map is locally invertible where the determinant of the
Jacobian matrix
\begin{equation}
\det J =1-\frac{\partial y}{\partial \overline{z}}
\overline{\frac{\partial y}{\partial \overline{z}}} = 1-\left| S_2
\left( z \right) \right|^2%
\label{General detJ}
\end{equation}
is different from zero. The points where the Jacobian determinant
vanishes are arranged in smooth closed curves called critical
curves. The images of these points through the lens map (\ref{Lens
equation}) in the source plane are called caustics. When a source
crosses a caustic, creation or destruction of pairs of images
occurs and the magnification diverges (Schneider, Ehlers \& Falco
1992).

This is all we need to start our search for secondary caustics in
close multiple systems. The fundamental hypothesis we make is
\begin{equation}
\left|z_i \right| \ll \sqrt{M} \; \; \; \forall i, %
\label{Perturbative hypothesis}
\end{equation}
where $M$ is the total mass of the system. In this way, the
distances between pairs of lenses will be very small with respect
to the Einstein radius of the lens that we would have if all the
masses were concentrated at the origin. This Einstein radius is
$\sqrt{M}$ in our notation. The relation (\ref{Perturbative
hypothesis}) allows us to consider the $z_i$'s as perturbative
parameters in a series expansion. Then we can solve the equation
$\det J =0$ at each order, writing its solutions as series
expansions in powers of the perturbative parameters. In this way
we shall find the critical curves of this system and study their
properties analytically.

\section{Number and positions of secondary critical curves}

Close multiple lenses have two classes of critical curves: the
main critical curve, resulting from the deformation of the
Einstein ring of the total mass lens, and the secondary critical
curves, forming inside the distribution of the masses.

Effectively, if we multiply the equation $\det J=0$ by the
quantity $\prod\limits_{i=1}^n \left| z- z_i \right|^4$, we get a
complex equation in $z$ and $\overline{z}$:
\begin{equation}
\prod\limits_{i=1}^n \left| z- z_i \right|^4- \left|
\sum\limits_{i=1}^n m_i \prod\limits_{j \neq i} \left(z-z_j
\right)^2 \right|^2=0. %
\label{Eq detJ}
\end{equation}

At the zero order, putting all $z_i$'s to zero, this equation
becomes
\begin{equation}
\left|z \right|^{4n-4} \left( \left|z \right|^4 - M^2 \right)=0.
\end{equation}
This equation has the solution $\left| z \right|=\sqrt{M}$, that
is the Einstein ring of the total mass lens. Taking this solution
as the starting point of a perturbative expansion, we get the main
caustic. The details of this calculation are in (Bozza 2000). But
the presence of the solution $z=0$ indicates that also this value
can be taken as the starting point for another expansion. This is
just the value we shall take to find the secondary critical
curves.

Having observed the zero order situation, we can start our
perturbative approach, searching for the first order solution.
Then we write the solution $z$ as a series expansion:
\begin{equation}
z=z_0+o \left( \left|z_i \right| \right),
\end{equation}
where $z_0$ is of the first order in $\left|z_i \right|$. Stopping
at the first order, we put $z=z_0$ in Eq. (\ref{Eq detJ}). We see
that the first term becomes of order $4n$, while the second is of
order $4n-4$. Then the latter dominates the first and Eq. (\ref{Eq
detJ}) is equivalent to
\begin{equation}
\sum\limits_{i=1}^n m_i \prod\limits_{j \neq i} \left(z_0-z_j
\right)^2=0. %
\label{Eq pos}
\end{equation}

This is a polynomial equation of degree $2n-2$. Then, for a system
of n close lenses, there are, at most, $2n-2$ points where the
Jacobian determinant vanishes (at the first order in $z_i$),
corresponding to $2n-2$ secondary critical curves.
This is the first main result of our work. It is
consistent with the binary lens, since two secondary critical
curves are predicted by this formula.

Eq. (\ref{Eq pos}) can be solved analytically for two and three
lenses, otherwise we have to resort to simple numerical methods.
In Sect. 6, we shall specify these and the following results for
the binary lens where a manageable expression for the positions of
the critical curves is available. For the triple lens, the
analytical solutions are too cumbersome to allow a detailed study.

Now, we have a straightforward way to calculate the positions of
the secondary critical curves for an arbitrary configuration of close
multiple lenses. Then, we can avoid the traditional blind
sampling of the Jacobian determinant on the lens plane and
reach, by this new method,
the full efficiency.

We take the generical solution $z_0$ of Eq. (\ref{Eq pos}) as the
first order term of our expansion. From now on, we use the
notation
\begin{equation}
S_k^0=S_k \left( z_0 \right).
\end{equation}
As both $z_0$ and $z_i$ are of the first order, according to our
perturbative expansion, $S_k^0$ has all denominators of order $k$
and then it is of order $-k$.

To continue our study we do not need an analytical expression for
$z_0$. We shall just use the fact that $z_0$ is a solution of Eq.
(\ref{Eq pos}), that is equivalent to say that
\begin{equation}
S_2^0=0.%
\label{S20}
\end{equation}

Of course, we have to distinguish between simple roots of Eq.
(\ref{Eq pos}) and roots of higher multiplicity. Remembering that
the $k^{\mathrm{th}}$ derivative of $S_2 \left( z\right)$ is
proportional to $S_{k+2} \left(z \right)$, we have the equivalence
between the following statements:
\begin{equation}
z_0\text{ is a root of multiplicity }p \Leftrightarrow S_{k+2}^0=0
\; \;  \forall k<p. %
\label{Mult equivalence}
\end{equation}

We shall treat separately the caustics coming from simple roots
(hereafter called simple caustics) and the caustics coming from
multiple roots (hereafter multiple caustics).

\section{Simple caustics}

These caustics are largely the most common as we explain in the
next section. So they surely have the most practical interest.

\subsection{Shape of the critical curves}

Once found the positions of the critical curves, we can carry on
our perturbative expansion to discover the shape of these curves.
So we put
\begin{equation}
z=z_0+\epsilon_2+\epsilon_3+\ldots,
\end{equation}
where $z_0$ is the position of one simple critical curve, found by
Eq. (\ref{Eq pos}), and $\epsilon_j$ are the corrections of order
$\left|z_i\right|^j$. It is convenient to use the original
equation $\det J=0$, which can be written in the form
\begin{equation}
1-S_2 \left(z_0+\epsilon_2+\epsilon_3 \right) \overline{S}_2
\left(\overline{z}_0+\overline{\epsilon}_2+\overline{\epsilon}_3 \right)=0,%
\label{detJ=0}
\end{equation}
starting from Eq. (\ref{General detJ})

The expansion of $S_2$ is
\begin{eqnarray}
& S_2\left(z_0+\epsilon_2+\epsilon_3 \right)=&S_2^0+ \nonumber \\%
& & -2 \epsilon_2 S_3^0+ \nonumber \\%
& & -2 \epsilon_3 S_3^0+3 \epsilon_2^2 S_4^0+ \ldots.
\end{eqnarray}
The first row is the order $-2$ and is null according to Eq.
(\ref{S20}). The second row is the order $-1$ and the third row is
the order zero. Inserting this expansion in (\ref{detJ=0}), the
lowest order equation is of order $-2$:
\begin{equation}
-4\left| \epsilon_2 \right|^2 \left| S_3^0 \right|^2=0,
\end{equation}
Being $z_0$ a simple root, $S_3^0 \neq 0$, so that $\epsilon_2=0$.

The successive terms in the expansion of the equation
(\ref{detJ=0}) are of order zero:
\begin{equation}
1-4\left| \epsilon_3 \right|^2 \left| S_3^0 \right|^2=0.
\end{equation}
From this equation we have
\begin{equation}
\left| \epsilon_3 \right|=\frac{1}{2 \left| S_3^0 \right|}.
\label{Critical r eq}
\end{equation}
Then the third order contains the first information on the shape
of the critical curve. Eq. (\ref{Critical r eq}) tells us that the
critical curve at position $z_0$ is a circle centered on $z_0$
with radius
\begin{equation}
r=\frac{1}{2 \left| S_3^0 \right|}. %
\label{Critical r}
\end{equation}
From the form of $S_3^0$ (see Eq. (\ref{Sk})), we see that the
closer the critical curve is to some mass, the higher the value of
$S_3^0$, the smaller the radius of the circle.

If we multiply all masses by a factor $\lambda$, the positions of
the critical curves do not change, because $\lambda$ factors out
from Eq. (\ref{Eq pos}), but their radii change as $\lambda^{-1}$.
If we do the same with the positions of the masses instead, the
positions of the critical curves scale as $\lambda$ and their
radii scale as $\lambda^3$.

\subsection{Caustics}

To find the caustics corresponding to the simple critical curves,
we just have to put the critical curve, in its obvious
parameterization
\begin{equation}
z\left( \theta \right)=z_0+r e^{i\theta} \; \; \; 0\leq \theta < 2
\pi,
\end{equation}
into the lens equation (\ref{Lens equation}) and expand to the
third order:
\begin{equation}
y\left( \theta \right)=-\overline{S}_1^0+z_0+\left( r
e^{i\theta}-\frac{e^{-2i\theta}}{S_3^0} \right). %
\label{Simple caustic}
\end{equation}

We can observe that the lowest order is $-1$ and is independent on
$\theta$. It represents the position of the caustic. From the
order of this term, we can deduce that these caustics can lie very
far from the origin of our system, going to infinity as the
distances among the masses are reduced to zero. The successive
term is $z_0$, which is of the first order and represents a
correction to the position. Finally, the shape of the caustic is
given by the third order.

The cusps of a caustic are characterized by the vanishing of the
tangent vector. To find them, we have to require that
\begin{equation}
\frac{\mathrm{d} y \left( \theta \right)}{\mathrm{d} \theta}=0
\end{equation}
and solve for $\theta$. Taking $y \left( \theta \right)$ from Eq.
(\ref{Simple caustic}), this equation can be simplified into
\begin{equation}
e^{3i\theta}+\sqrt{\frac{\overline{S}_3^0}{S_3^0}}=0,
\end{equation}
whose solutions are
\begin{equation}
\theta_k=-\frac{1}{3}\arg \left( S_3^0 \right) +\frac{2k \pi}{3}
\; \; k=0,1,2,
\end{equation}
where $\arg$ yields the argument of a complex number.

We have three cusps. So, in any close multiple system, having only
simple secondary caustics, these caustics have a triangular shape.

Finally, we calculate the area of these caustics. This can be done
by the integral
\begin{equation}
A=\int\limits_\gamma y_2 \mathrm{d} y_1,
\end{equation}
where $\gamma$ is the caustic in its clockwise direction. We have
\begin{equation}
A=-\frac{1}{4i}\int\limits_0^{2\pi} \left[y \left( \theta \right)
-\overline{y}\left( \theta \right) \right]\partial_\theta \left[y
\left( \theta \right) +\overline{y}\left( \theta \right) \right]
\mathrm{d} \theta.
\end{equation}
The minus in the right member comes from the fact that our
parameterization is counterclockwise. The integral only involves
complex exponential functions and the result is
\begin{equation}
A=\frac{1}{2}\pi r^2.
\end{equation}
So the extension of the simple caustics is of the sixth order in
the separations among the lenses, justifying the evasive nature of
these caustics.

\begin{figure}
 \resizebox{\hsize}{!}{\includegraphics{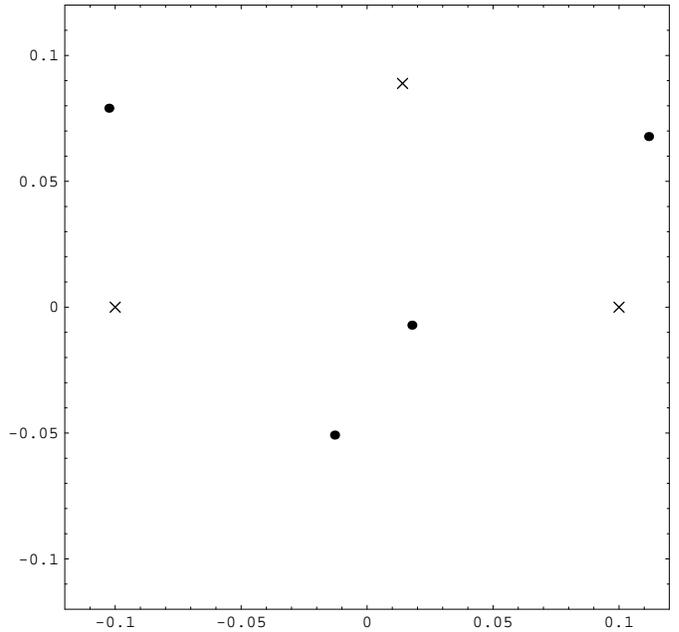}}
 \caption{Positions of the four secondary critical curves %
 (indicated by the small dots) for a triple system with %
 masses $m_1=0.25$, $m_2=0.25$, $m_3=0.5$ and positions %
 $z_1=0.1$, $z_2=-0.1$, $z_3=0.014+i0.089$ (indicated by %
 the three crosses).}
 \label{Fig simple critical curves}
\end{figure}

\begin{figure*}
 \resizebox{12cm}{!}{\includegraphics{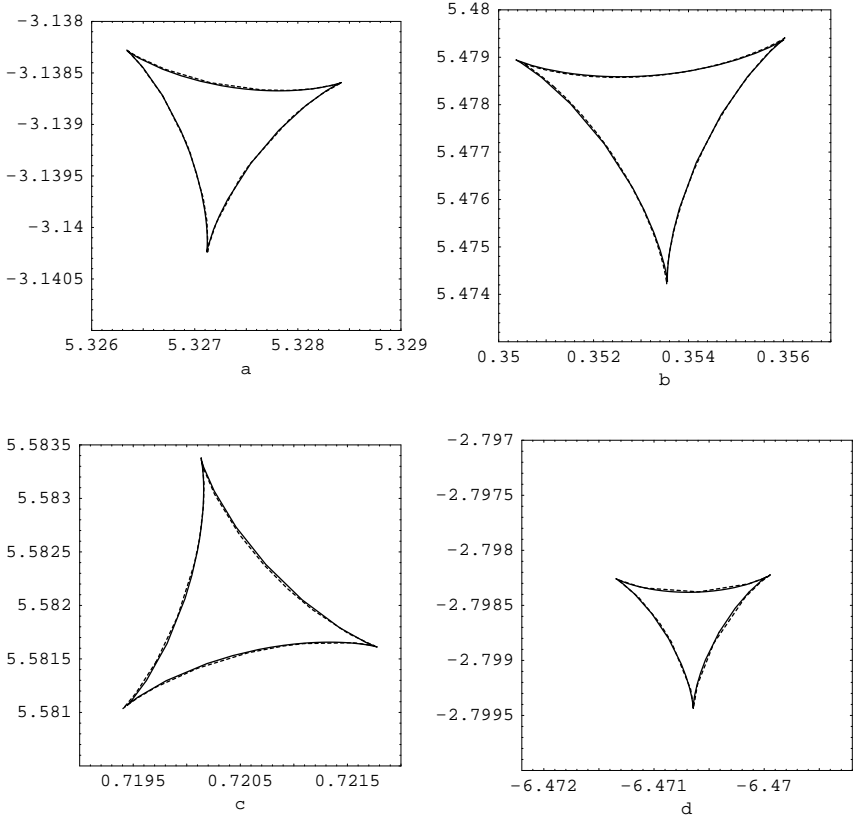}}
  \hfill
  \parbox[b]{55mm}{
 \caption{Here are the four caustics produced by the critical curves %
 of Fig. \ref{Fig simple critical curves} from left to right. The solid %
 curve is the perturbative caustic and the dashed line is the numerical
 one.}}
 \label{Fig simple caustics}
\end{figure*}

With our expansions, we have attained considerable analytical information
about the secondary caustics establishing their shape, the area, the
number of cusps in a completely general way. However, since these
results are the fruit of perturbative approximations, it is important
to discuss their accuracy.
So we propose a comparison between our perturbative results and
the numerical ones in a typical situation.
We consider a system constituted by three
lenses disposed as in Fig. \ref{Fig simple critical curves}.
According to our previous statement, this system can form, at
most, four simple secondary critical curves. For our choice of
parameters, we display their positions in the same figure. The
caustics produced by these curves are shown in Fig. 2 where they
are compared to the numerical ones. We have taken the distances
among the masses of this distribution to be one tenth of the total
Einstein radius. Even for this not too small value, the positions
and the shapes of the secondary caustics are reproduced with a
striking accuracy. It is also to be noted that the quality of
numerical results is improved thanks to the guide provided by
perturbative results.

So we see that the
analytical formulae derived in this section are very good
approximations to the quantitative characteristics of the
secondary caustics, proving to be highly reliable.

\section{Multiple caustics}

In this section, we consider the case where $z_0$ is a multiple
root of Eq. (\ref{Eq pos}). The parameters space of a system with
n lenses is $3n-4$ dimensional, since each mass adds three
parameters (its mass and its coordinates in the lens plane). Four
parameters can be eliminated by considering equivalent those
systems differing by a global translation or rotation and/or by a
global scale factor. Thus, for example, the binary lens is
completely characterized by the mass ratio and the separation
between the lenses.

The requirement of a double root in Eq. (\ref{Eq pos}) translates
into the vanishing of the derivative of this equation with respect
to $z$. This is one constraint equation, then the points of the
parameters space producing multiple roots constitute a $3n-5$
dimensional hypersurface, thus having measure zero. For this
reason, the occurrence of multiple roots is relatively rare.
Anyway, very interesting features emerge, justifying a detailed
study of these particular cases.

\subsection{Critical curves}

Suppose that $z_0$ is a root with multiplicity $p$. We have to
find the correct order of the perturbation to insert in the
equation $\det J =0$, representing the shape of our critical
curve. According to the equivalence (\ref{Mult equivalence}), the
$S_{k+2}^0$ with $k<p$ are null. Then, we put
\begin{equation}
z=z_0+\epsilon,
\end{equation}
with the order (that we shall indicate by  $q$) of $\epsilon$ to
be found. We only assume that $q$ be higher than one. Then, the
expansion of $S_2 \left( z \right)$ is
\begin{multline}
S_2\left( z_0+\epsilon \right)=S_2^0-2 \epsilon S_3^0+3 \epsilon
S_4^0+ \ldots \\ \ldots +\left(-1\right)^k \left( k+1 \right)
\epsilon^k S_{k+2}^0+ \ldots.
\end{multline}

The $k^\mathrm{th}$ term is of order $q k-\left(k+2\right)$ and
the first term to be non--null is that for $k=p$. When we put this
expansion in the equation $\det J=0 $, the first non--null term is
\begin{equation}
-\left( p+1 \right)^2 \left|\epsilon^{p}S_{p+2}\right|^2
\end{equation}
having order $2q p-2\left(p+2\right)$. If this order is less than
zero, we just get from $\det J=0$ that $\epsilon=0$, but if the
order of this term is zero, then the zero order expansion of $\det
J=0$ also involves another term (equal to $1$):
\begin{equation}
1-\left( p+1 \right)^2 \left|\epsilon^{p}S_{p+2}\right|^2=0
\end{equation}
and the equation gives the non--trivial solution
\begin{equation}
\left|\epsilon \right|=\frac{1}{\left[\left( p+1 \right)
\left|S_{p+2}\right|\right]^{1/p}}.
\end{equation}
This happens when the order of $\epsilon$ is $q=\frac{2+p}{p}$.
This is consistent with the result of the previous section,
because, for $p=1$, $q=3$. For $p=2$, we have that the first non
trivial order is the second and, for $p=3$, it is the order
$\frac{5}{3}$. When $p$ increases, the order of this perturbation
decreases, approaching 1 as a limit. This means that at high
multiplicities, the perturbative expansion becomes always less
accurate, requiring ever more terms for an adequate description of
the caustics. Anyway, the main characteristics of the caustics can
be derived retaining just the first correction and that is what we
shall do.

The critical curve just derived is again a circle with radius
\begin{equation}
r=\frac{1}{\left[\left( p+1 \right)
\left|S_{p+2}\right|\right]^{1/p}}, %
\label{Mult critical r}
\end{equation}
becoming greater with the multiplicity.

\subsection{Caustics}

We take, as before, the parameterization
\begin{equation}
z \left(\theta \right)=z_0+ r e^{i\theta}
\end{equation}
for the critical curve, with $r$ given by Eq. (\ref{Mult critical
r}). Putting this expression into the lens equation (\ref{Lens
equation}) and expanding to the $q^{\mathrm{th}}$ order, we get
\begin{equation}
y\left( \theta \right)=-\overline{S}_1^0+z_0+r \left[
e^{i\theta}+\left( -1 \right)^p \frac{e^{-\left( p+1 \right)
i\theta}}{ p+1 } \sqrt{\frac{\overline{S}_{p+2}^0}{ S_{p+2}^0}}
\right].%
\label{Mult caustic}
\end{equation}

\begin{figure}
 \resizebox{\hsize}{!}{\includegraphics{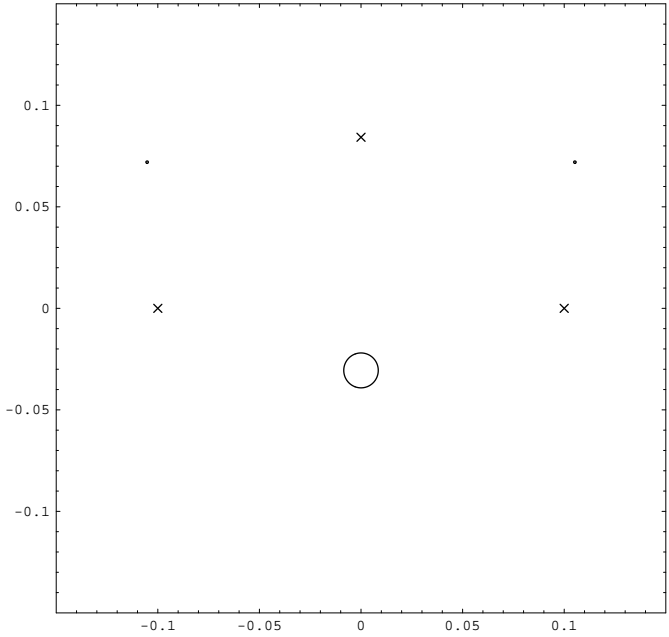}}
 \caption{Critical curves in the event of a double root. %
  The masses are indicated by the crosses. The double critical %
  curve is the one at the bottom--center, while the other two %
  are on the top--left and top--right and are hardly visible in %
  this picture.}
 \label{Fig Double critical curve}
\end{figure}

The $q^{\mathrm{th}}$ order is the first depending on $\theta$ and
determines the shape of the caustic.

To understand this shape, we calculate the cusps as in the
previous section. The equation for the cusps is
\begin{equation}
e^{\left(p+2 \right) i\theta}+\left( -1 \right)^{p+1}
\sqrt{\frac{\overline{S}_{p+2}^0}{S_{p+2}^0}}=0
\end{equation}
and its solutions are
\begin{equation}
\theta_k=\frac{\left(-1 \right)^p}{p+2 }\arg \left(
S_{p+2}^0\right) +\frac{2k \pi}{p+2} \; \; \; 0 \leq k<p+2.
\end{equation}

Now we have $p+2$ cusps. This is a very interesting result,
because the caustic assumes the shape of a regular polygon with
$p+2$ curved sides.

The area of the multiple caustic can be calculated in the same way
as for the simple one. We just give the result:
\begin{equation}
A= \frac{p}{p+1}\pi r^2.
\end{equation}
It is of order $2\frac{2+p}{p}$. So, increasing the multiplicity
from 1 to infinity, the order of the area lowers from 6 to 2 and
the extension of the caustic becomes ever more important.

Some other consideration about the limit for $p\rightarrow \infty$
can be done. The number of cusps become infinite and, from Eq.
(\ref{Mult caustic}), we see that the caustic becomes a circle of
radius $r$. In fact, the area becomes $\pi r^2$.

\subsection{An example: a double caustic in a triple lens}

\begin{figure}
 \resizebox{\hsize}{!}{\includegraphics{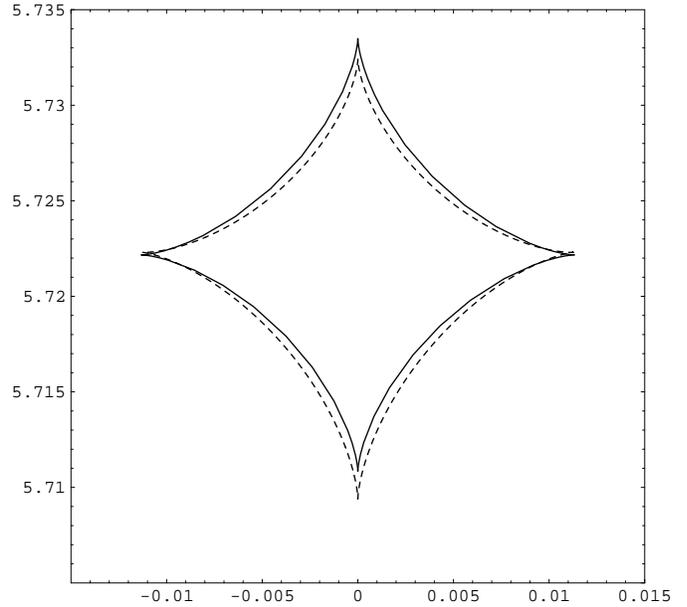}}
 \caption{Caustic corresponding to the double critical curve %
 of Fig. \ref{Fig Double critical curve}. The dashed curve is %
 the numerical caustic and the solid line is the perturbative %
 one.}
 \label{Fig Double caustic}
\end{figure}

Now we shall practically see how our formulae work in the case of
a multiple caustic. We consider three masses: $m_1=0.25$,
$m_2=0.25$ and $m_3=0.5$. We fix the positions of the first two:
$z_1=0.1$, $z_2=-0.1$; but we let the third free for the moment.
We simultaneously solve Eq. (\ref{Eq pos}) and its derivative with
respect to $z_0$, for the two unknowns $z_0$ and $z_3$. We find
six possible positions of the third mass, giving rise to a double
root of the positions equation. None of them is a triple root. Two
of these positions are on the $x_2$-axis. We choose one of them:
$z_3=i 0.084263$. The double root is in $z_0=-i0.0299$. In Fig.
\ref{Fig Double critical curve}, we see that the critical curve in
this point is much greater than the other two. In fact, the radius
of the double critical curve, calculated by Eq. (\ref{Mult
critical r}), is $8.49 \times 10^{-3}$, while the radius of the
two simple critical curves is $6.13 \times 10^{-4}$, according to
Eq. (\ref{Critical r}).

In Fig. \ref{Fig Double caustic}, we show the caustic generated by
this double critical curve. The geometry is correctly predicted by
our perturbative expansion: there are four cusps in a double
caustic. We see that the approximation is less accurate than
before, as we anticipated in our discussion about the order of the
perturbation. However, for double caustics, it is not so difficult
to add another term to the perturbative expansion and reach the
same accuracy of the simple caustics. The third order term in the
critical curve depends on $\theta$:
\begin{equation}
\epsilon_3=6 r^2 \mathrm{Re}\left[ \overline{S}_4^0 S_5^0 e^{i
\theta} \right].
\end{equation}
The successive term in the caustic is
\begin{equation}
e^{i \theta} \epsilon_3+3 e^{-3i \theta} r^2 \epsilon_3
\overline{S}_4^0- r^4 e^{-4i \theta} \overline{S}_5^0.
\end{equation}

Double caustics, and, more generally, multiple caustics, are
formed by the union of small caustics, in some sense. Another
interesting question is: what happens if we change the parameters
in the neighbourhood of our particular choice producing the double
root? We expect the double critical curve to separate into two
smaller ovals and the double quadrangular caustic to break into
two triangular ones; but this can happen in different ways.

\begin{figure*}
 \resizebox{15cm}{!}{\includegraphics{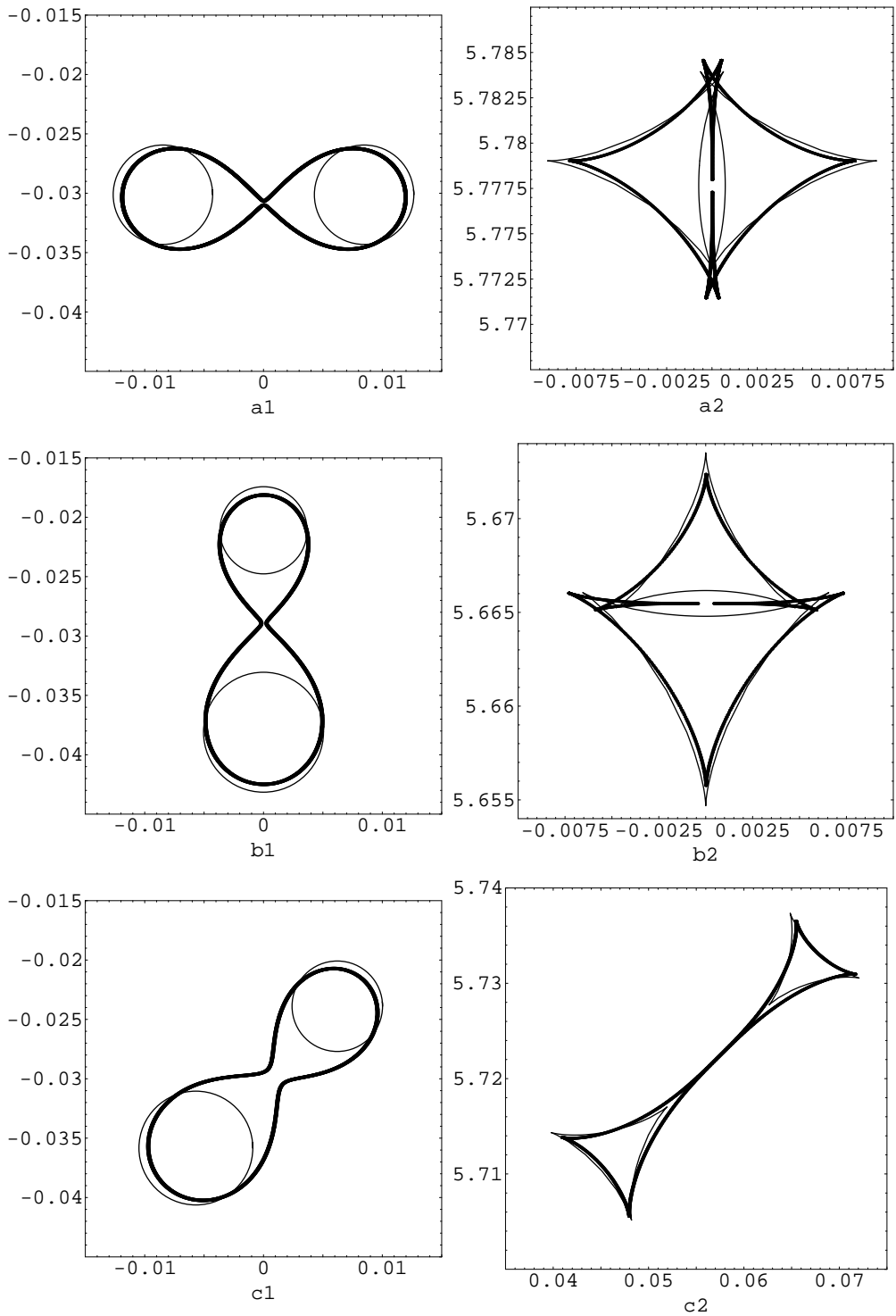}}
 \hfill
 \parbox[b]{15cm}{
 \caption{Critical curves and caustics for $z_3$ close to %
 $i 0.084263$. The left column shows the critical curves and %
 the right column the caustics. The thick lines are the numerical %
 curves and the thin lines are the perturbative ones. The choice of %
 $z_3$ in cases $a$, $b$ and $c$ are given in the text.}
 \label{Fig Double neighbourhood}}
\end{figure*}

In this regime, the perturbative caustics are simple.
However, as the parameters tend to give the
double root, $S_3^0$ tends to zero, yielding a diverging $r$ for
the simple critical curves, according to Eq. (\ref{Critical r}).
The transition with the formation of the double critical curve is
thus not reproduced.
Guided by perturbative approximations, the break of the double
caustic, when $z_3$ moves out from the position $i
0.084263$, can be investigated numerically. The results are shown in
Fig. \ref{Fig Double neighbourhood}.

In case $a$, $z_3=i 0.082787$, i.e. we have moved the third mass
towards the others. The critical curve breaks in the horizontal
direction. Looking just at the thick line in Fig. \ref{Fig Double
neighbourhood}a2, representing the numerical caustic, we see that
the top cusp and the bottom cusp develop a butterfly geometry. At
some critical value, these butterflies touch and the two resulting
triangular caustics move away along the horizontal direction. We
have displayed in the same plot the perturbative caustics too.
Obviously, they are simple caustics, so they cannot show the
butterfly geometry but they can help in understanding how the
separation occurs. We also notice that the simple caustics
cover the area of the
numerical transition double caustic very well constituting a significant
approximation anyway.

In case $b$, $z_3=i 0.085759$, so that the third mass is farther
from the others. Now the critical curve breaks in the vertical
direction and so does the caustic. The left and the right cusps
transform into butterflies. These butterflies are slightly
distorted by the fact that the resulting simple caustics have
different sizes: the one on the top is smaller than the other.

In case $c$, $z_3=0.00147+i 0.084263$. We have displaced the third
mass in the horizontal direction. The critical curve breaks
diagonally and so does the caustic. But this time the transition
occurs with a simple beak--to--beak singularity rather than with
butterflies. While in the previous situations the two simple
caustics in the last step of the separation touch with a fold,
here they touch with a cusp.

\section{Secondary caustics in binary lensing}

In this section, we specify our results for the binary case where
simple analytical formulae can be written. Let's consider two
masses placed on the horizontal axis and let's choose the origin
in the centre of mass. We call the separation between the masses
$a$, then we have $z_1=\frac{m_2 a}{m_1+m_2}$ and $z_2=-\frac{m_1
a}{m_1+m_2}$. Eq. (\ref{Eq pos}) is of second degree. Its
solutions are
\begin{equation}
z_0=a\frac{m_2-m_1 \pm i \sqrt{m_1 m_2}}{m_1+m_2}.
\end{equation}
They are always simple and lie on a circle of radius $a/2$
centered in the middle of the two masses.

The radius of the two ovals is the same:
\begin{equation}
r=\frac{\sqrt{m_1 m_2} a^3}{2 \left( m_1+m_2 \right)^2}.
\end{equation}
Its maximum value $\frac{a^3}{4 M_\mathrm{tot}}$ is reached when
the two masses are equal, in fact, in this case, their distance
from the two masses is maximum.

The two caustics are given by the following expression:
\begin{multline}
y \left( \theta \right)=\frac{m_1-m_2}{a} \mp i \frac{2 \sqrt{m_1
m_2}}{a}+ \\ %
+a\frac{m_2-m_1}{m_1+m_2} \pm i a \frac{\sqrt{m_1
m_2}}{m_1+m_2} +\\ %
+ a^3 \sqrt{m_1 m_2}\left[ \frac{  e^{i \theta}}{2 \left( m_1+m_2
\right)^2} \pm \frac{i e^{-2 i \theta}}{4 \left( \sqrt{m_1} \pm i
\sqrt{m_2}\right)^4} \right].
\end{multline}

Their cusps are at positions
\begin{equation}
\theta_k=- \arg \left[\pm i \left(\sqrt{m_1} \pm i \sqrt{m_2}
\right)^4 \right] +\frac{2k\pi}{3} \; \; k=0,1,2.
\end{equation}

Their area is
\begin{equation}
A=\frac{\pi m_1 m_2 a^6}{8 \left( m_1+m_2 \right)^4},
\end{equation}
reaching the maximum value $\frac{\pi a^6}{32M^2_{\mathrm{tot}}}$
in the equal masses case.

\section{Summary}

Multiple Schwarzschild lensing represents a very rich terrain for
the exploration of caustics in gravitational lensing. The
occurrence of different kinds of singularities stimulates new
investigations.

In this paper we have applied perturbative methods to secondary
caustics,
forming when the masses are close each other with respect to the
total Einstein radius. In this way we have been able to establish
the number of the caustics for any lens configuration, the positions
and the shapes, with a complete characterization of the geometries
arising in all cases. Moreover, quantitative formulae for the area
and other features of these objects have been given.
As we have seen, in the
most common case, the shape of the
simple caustics is always triangular. Anyway, multiple caustics exist
developing a great
variety of behaviours, giving rise, to curves having a number of cusps
ranging from four to
infinity. The breaking of multiple caustics can follow different
ways depending on how the parameters of the system change.

\begin{acknowledgements}
I would like to thank Gaetano Scarpetta and Salvatore Capozziello
for their helpful comments on the manuscript.

Work supported by fund ex 60\% D.P.R. 382/80.
\end{acknowledgements}


\begin{thebibliography}{}

\bibitem{Bozza a}  Bozza V., 1999, A\&A 348, 311

\bibitem{Bozza b}  Bozza V., 2000, A\&A in press, astro-ph/9910535

\bibitem{Chang & Refsdal 1}  Chang K., Refsdal S., 1979, Nat 282, 561

\bibitem{Chang & Refsdal 2}  Chang K., Refsdal S., 1984, A\&A 132, 168

\bibitem{Dominik}  Dominik M., 1999, A\&A 349, 108

\bibitem{Erdl & Schneider}  Erdl H., Schneider P., 1993, A\&A 268, 453

\bibitem{Gaudi et al.}  Gaudi B.S., Naber R.M., Richard M., Sackett P.D.,
1998, ApJ 502, 33

\bibitem{Kayser et al.}  Kayser R., Wei{\ss} A., Refsdal S.,
Schneider P., 1988, A\&A 214, 4

\bibitem{Schneider, Ehlers & Falco}  Schneider P., Ehlers J., Falco
E.E.,1992, Gravitational lenses. Springer-Verlag, Berlin

\bibitem{Schneider & Weiss}  Schneider P., Wei{\ss} A., 1986, A\&A 164, 237

\bibitem{Schramm et al.}  Schramm T., Kayser R., Chang K. et al.,
1993, A\&A 268, 350

\bibitem{Witt}  Witt H.J., 1990, A\&A 236, 311

\bibitem{Witt & Petters}  Witt H.J., Petters A.O., 1993, J. Math. Phys.
34, 4093

\bibitem{Zheng & Gould} Zheng Z., Gould A., 2000, submitted to
ApJ, astro-ph/0001199

\end{thebibliography}
\end{document}